\documentclass[noshowpacs,pra,aps,superscriptaddress,floatfix]{revtex4}
\usepackage{amsmath}
\usepackage{amssymb,mathrsfs}
\usepackage{graphicx}
\usepackage[dvipsnames]{xcolor}

\usepackage[utf8]{inputenc}

\newcommand{\prt}{\partial}
\newcommand{\om}{\omega}

\newcommand{\ka}{\kappa}
\newcommand{\vphi}{\varphi}

\newcommand{\ou}{\overline{u}}
\newcommand{\bu}{\mathbf{u}}
\newcommand{\orho}{\overline{\rho}}

\begin{document}
\title{Hamilton theory of NLS equation soliton motion}

\author{A. M. Kamchatnov}
\affiliation{Institute of Spectroscopy, Russian Academy of Sciences,
Moscow, Troitsk, Russia}
\affiliation{Skolkovo Institute of Science and Technology, Skolkovo,
Moscow Region, Russia}

\begin{abstract}
We suggest the method of derivation of Hamilton equations which describe the motion
of solitons along non-uniform and time dependent large-scale background in case
of wave dynamics described by the completely integrable equations in the
Ablowitz-Kaup-Newell-Segur scheme. The method is based on development of old
Stokes' argumentation which allows one to continue analytically some relationships
derived for linear waves to the soliton region. It is presented here for a particular
case of the defocusing nonlinear Schr\"{o}dinger equation. We formulate the condition
when the external potential should only be taken into account for the background evolution,
and in this case we obtain the Newton equation for the soliton dynamics.
\end{abstract}



\maketitle

\section{Introduction}\label{intro}

Propagating without change of form localized nonlinear waves called solitons are
often likened to particles, especially when their dynamics is described by the
completely integrable equations, since in this case they interact elastically with
each other (see, e.g., Refs.~\cite{nmpz-80,as-81,newell-85}). This analogy between
solitons and particles still remains reasonable when solitons propagate in external
fields or along weakly non-uniform and slowly evolving background provided one can
define with good enough accuracy the time-dependent soliton's coordinate $x=x(t)$.
In many such situations, the dynamics of the variable $x=x(t)$ coincides with the
dynamics of a point particle if one neglects small corrections of the order of
magnitude of the ratio of soliton's width to the characteristic size at which
the external field or the background change (see, e.g., \cite{km-89,kosevich-90}).
However, even in this case the soliton's motion can cause a counterflow in the
surrounding background and such a counterflow can make a considerable contribution
to the soliton's momentum in spite of small amplitude of the counterflow wave
\cite{km-77,km-78,kn-80}. Especially spectacular the counterflow effects are
demonstrated in the motion of dark solitons in Bose-Einstein condensate with
repulsive interaction between atoms when the condensate's dynamics is described by
the Gross-Pitaevskii equation \cite{gross-61,pit-61} which coincides mathematically
with the defocusing nonlinear Schr\"{o}dinger (NLS) equation. In this case,
excitation of a dark soliton is accompanied by formation of the phase jump across
the soliton, but since the condensate wave function must remain single valued,
the counterflow appears around the soliton which compensates the soliton's phase jump.
As a result, the actual soliton's dynamics with account of the counterflow differs
drastically from the dynamics which could be expected for situations without the
counterflow. For example, if the condensate is confined in a harmonic trap with
the frequency $\om_0$, then its center of mass oscillates after excitation of a soliton
with the same frequency $\om_0$, whereas the soliton's coordinate $x(t)$ oscillates
with the frequency $\om_0/\sqrt{2}$, that is the counterflow deforms essentially
the background flow caused by the soliton's motion. Thus, the soliton's dynamics
must be developed with account of the counterflow which makes essential contributions
to the dynamical variables of such nonlinear excitations.

Hamilton theory of the Korteweg-de Vries (KdV) equation was developed in
Ref.~\cite{zf-71} and then it was extended to a wide class of completely integrable
equations (see, e.g., Ref.~\cite{dickey-03} and references therein), however in this
theory dynamics reduces to interaction of solitons with each other and with harmonic
waves of linear spectrum, so that in this approach there is no any approximations
distinguishing narrow solitons from a smooth wide and slowly evolving background.
An approximate Hamiltonian
dynamics of KdV solitons propagating along a smooth evolving background was built in
Ref.~\cite{mts-79} by means of a quite complicated technique of generalized functions.
In the Gross-Pitaevskii (NLS equation) theory, the canonical momentum and energy of
a dark soliton in a uniform condensate were calculated in Ref.~\cite{shevchenko-88}
(see also Refs.~\cite{ps-03,pit-16}) by means of physical reasoning and generalization
of this theory to non-uniform condensate made in Ref.~\cite{ik-22} allowed one to
reproduce known results \cite{ba-2000,kp-04,kk-10} about dark soliton's dynamics and
to obtain some new ones. However, a simple and general enough Hamiltonian approach to
soliton's dynamics along a non-uniform evolving background had not been developed
until recently, when such an approach was suggested in Ref.~\cite{ks-23a} for a
generalized KdV equation. In particular case of the completely integrable KdV
equation this approach reproduced very simply the results of Ref.~\cite{mts-79}.
The aim of this article is to extend this approach to nonlinear wave equations
completely integrable in framework of the Ablowitz-Kaup-Newell-Segur (AKNS)
scheme \cite{akns-74}. We apply the method to the defocusing NLS equation with
account of the external potential for situations when non-uniformity effects are
more essential then the direct contribution of the external potential to the
soliton's energy. Simplicity of the method suggests that it can be extended to
other integrable equations beyond the AKNS scheme.

\section{Hamilton equations for NLS dark soliton dynamics}

We take the NLS equation in standard dimensionless notation
\begin{equation}\label{eq1}
	i\psi_t+\frac12\psi_{xx}-|\psi|^2\psi=U(x)\psi,
\end{equation}
where $U(x)$ is an external potential. It is convenient to transform this equation
by means of the substitution
\begin{equation}\label{eq2}
	\psi(x,t)=\sqrt{\rho(x,t)}\exp\left({i}\int^x u(x',t)dx'\right),
\end{equation}
to the system
\begin{equation}\label{eq3}
	\begin{split}
		&\rho_t+(\rho u)_x=0,\\
		&u_t+uu_x+\rho_x+\left(\frac{\rho_x^2}{8\rho^2}
		-\frac{ \rho_{xx}}{4\rho}\right)_x= -U_x.
	\end{split}
\end{equation}
Then in the BEC context $\psi$ is the condensate's wave function, $\rho$ is its density
and $u$ its flow velocity. The system (\ref{eq3}) with $U=0$ has the soliton solution
\begin{equation}\label{eq4}
	\begin{split}
		\rho&=\overline{\rho}-\frac{\overline{\rho}-(V-\overline{u})^2}
{\cosh^2\left[\sqrt{\overline{\rho}-(V-\overline{u})^2}(x-Vt)\right]}, \\
		u&=V-\frac{\overline{\rho}(V-\overline{u})}{\rho},
	\end{split}
\end{equation}
where $\overline{\rho}$ and $\overline{u}$ denote, correspondingly, the density and
the flow velocity at infinity and $V$ is the soliton's velocity in a ``laboratory''
reference frame. The soliton's size is characterized by its inverse half-width
\begin{equation}\label{eq6}
  \ka=2\sqrt{\orho-(V-\ou)^2},
\end{equation}
so that at the soliton's tails we have $\orho-\rho\propto\exp[\pm\ka(x-Vt)]$. Consequently,
the velocity $V$ is related with $\ka$ by the formula
\begin{equation}\label{eq7}
  V=\ou+\sqrt{\orho-\frac{\ka^2}{4}},
\end{equation}
where we assume that the relative soliton's velocity $V-\ou$ is positive.

Another important characteristic of the system (\ref{eq3}) with $U=0$ is the dispersion
relation for linear waves propagating along a uniform background
$\rho=\orho$, $u=\ou$. Linearization of the system (\ref{eq3}) with respect to small
deviations $\rho',u'\propto\exp[i(kx-\om t)]$ of the variables $\rho,u$ from their background
values readily leads to the Bogoliubov law
\begin{equation}\label{eq8}
  \om=k\left(\ou+\sqrt{\orho+\frac{k^2}{4}}\right).
\end{equation}
As was noticed by Stokes \cite{stokes} in a similar situation for the KdV equation,
the soliton's velocity (\ref{eq7}) is expressed in
terms of the dispersion relation (\ref{eq8}) by the formula
\begin{equation}\label{eq9}
  V=\frac{\om(i\ka)}{ik}=\ou+\sqrt{\orho-\frac{\ka^2}{4}},
\end{equation}
which has clear physical meaning: both linear waves and soliton's tails obey the same
linearized equations, so that one solution transforms to the other by means of the
replacement  $k\leftrightarrow i\ka$. Since the tails move with the soliton's velocity,
the phase velocity of linear wave transforms then to the soliton's velocity.

Another expression for the velocity of a soliton moving through the non-uniform background
can be obtained in the following way. We assume that soliton's width $\sim\ka^{-1}$ is
always much smaller during its motion, than the characteristic length $l$, at which the
background variables change, so we can consider $\orho$ and $\ou$
constant at distances of order of magnitude $\sim\ka^{-1}$. As a result, we can separate
the large-scale evolution of the background $\orho=\orho(x,t),\ou=\ou(x,t)$ from
soliton's motion and assume that this evolution obeys the dispersionless equations
\begin{equation}\label{eq10}
	\begin{split}
		\orho_t+(\orho\,\ou)_x=0,\qquad
		\ou_t+\ou\,\ou_x+\orho_x=0,
	\end{split}
\end{equation}
which are obtained form Eqs.~(\ref{eq3}) by neglecting the terms with higher order derivatives
$\sim l^{-3}$. We also take here $U_x=0$ assuming that the gradient of the external potential
is small and for $\ka^{-1}\ll l$ it does not affect the shape of narrow solitons which energy and
momentum we wish to calculate. If the solution $\orho=\orho(x,t),\ou=\ou(x,t)$ of Eqs.~(\ref{eq10})
is known, then the instant value of soliton's velocity is obtained by substitution of $\orho,\ou$
into Eq.~(\ref{eq7}), where $\ka$ becomes also a function of time which changes along the soliton's
path. To find another relationship between $\ka=\ka(t)$ and the local values of the background
variables, we turn again to Stokes' argumentation and relate the motion of a narrow soliton with
the inverse half-width  $\ka\gg1/l$ along a non-uniform background with the propagation of a wave packet
with the carrier wave number $k\gg1/l$ along the same background. As is known (see, e.g.,
Ref.~\cite{LL-2,kyu-80}), propagation of such wave packets obeys the Hamilton equations
\begin{equation}\label{eq11}
\frac{d x}{d t} = \frac{\prt \omega }{\prt k},\qquad
\frac{d k}{d t} = -\frac{\prt \omega }{\prt x},
\end{equation}
where $x=x(t)$ is the packet's coordinate which can be defined with good enough accuracy due to
inequality $k\gg1/l$. The Hamiltonian $\om=\om(k,\orho,\ou)$ is given in our case by Eq.~(\ref{eq8})
and it depends on $x$ and $t$ via the background variables $\orho=\orho(x,t),\ou=\ou(x,t)$, which
are solutions of Eqs.~(\ref{eq10}). If we demand now that the wave number $k$ also depends on $x$
and $t$ only via the background variables, then we find \cite{sk-23} that such a function
$k=k(\orho,\ou)$ must satisfy equations
\begin{equation}\label{eq12}
  \frac{\prt k^2}{\prt\orho}=-4,\qquad \frac{\prt k^2}{\prt\ou}=-2\sqrt{k^2+4\orho},
\end{equation}
which have the solution
\begin{equation}\label{eq13}
  k^2=(q-\ou)^2-4\orho,
\end{equation}
where $q$ is an integration constant defined by a value of $k$ at some initial point in the
$(\orho,\ou)$-plane. It is worth noticing that existence of such a function is related with
complete integrability of the NLS equation (see \cite{kamch-23}), and this statement is confirmed
by other examples \cite{ks-23b}. Accordingly, formula (\ref{eq13}) and dispersion relation
(\ref{eq8}) turn out to be related with quasi-classical limit of the Lax pair of the equation
under consideration (see \cite{kamch-23,ks-23b}).

Now, following Stokes argumentation \cite{stokes}, we assume that the inverse half-width $\ka$
of a soliton moving along the same background is related with the local values of the variables
$\orho=\orho(x,t),\ou=\ou(x,t)$ by the formula
\begin{equation}\label{eq14}
  \ka^2=4\orho-(q-\ou)^2
\end{equation}
obtained from Eq.~(\ref{eq13}) by the replacement $k\to i\ka$. Here $q$ is also defined by the
soliton's parameters at the initial point of its path. Substitution of this relationship into
Eq.~(\ref{eq7}) gives (for $q-\ou>0$) a very simple expression the soliton's velocity:
\begin{equation}\label{eq15}
  \frac{dx}{dt}=V=\frac12(\ou(x,t)+q).
\end{equation}
Solution of this equation yields the soliton's path in situations when the external potential
is absent. We will use Eq.~(\ref{eq15}) for development of fuller Hamilton theory of soliton
motion with account of the external potential.

Calculations become simpler if we express the soliton's parameters $\ka,V$ and the background
variable $\ou$ at the point of soliton's position in the following form:
\begin{equation}\label{eq16}
  \ka=2\sqrt{\orho}\sin\vphi,\qquad V=\ou+\sqrt{\orho}\cos\vphi,
\end{equation}
\begin{equation}\label{eq17}
  \ou+2\sqrt{\orho}\cos\vphi=q=\mathrm{const},
\end{equation}
where $\vphi=\vphi(x,t)$ is a function of the space coordinate $x$ and time $t$.
Then Eq.~(\ref{eq14}) becomes an identity and differentiations of another identity (\ref{eq17})
along the soliton's path gives with account of Eq.~(\ref{eq16})
\begin{equation}\nonumber
  \begin{split}
  \ou_t+(\ou+\sqrt{\orho}\cos\vphi)\ou_x+
  \frac{1}{\sqrt{\orho}}[\orho_t+(\ou+\sqrt{\orho}\cos\vphi)]\cos\vphi-
  2\sqrt{\orho}\sin\vphi\frac{d\vphi}{dt}=0.
  \end{split}
\end{equation}
If we eliminate $\ou_t$ and $\orho_t$ with the help of dispersionless equations (\ref{eq10}),
then we obtain
\begin{equation}\label{eq19}
  \frac{d\vphi}{dt}=-\left(\sqrt{\orho}\right)_x\sin\vphi.
\end{equation}

Turning to derivation of expressions for the canonical momentum $p$ and Hamiltonian $H$ of a
soliton moving along the background $\orho=\orho(x,t),\ou=\ou(x,t)$, we interpret Eq.~(\ref{eq15})
as the Hamilton equation
\begin{equation}\label{eq20}
  \frac{dx}{dt}=\ou+\sqrt{\orho}\cos\vphi=\left.\frac{\prt H}{\prt p}\right|_x.
\end{equation}
Its integration gives
\begin{equation}\label{eq21}
  H=\ou p+\orho^{3/2}\int^{\vphi}\cos\vphi\cdot f'(\vphi)d\vphi,
\end{equation}
where we assumed that the canonical momentum can be expressed in the form
\begin{equation}\label{eq22}
  p=\orho f(\vphi).
\end{equation}
To find the function $f(\vphi)$, we use the second Hamilton equation
\begin{equation}\label{eq23}
  \frac{dp}{dt}=-\left.\frac{\prt H}{\prt x}\right|_p,
\end{equation}
that is
\begin{equation}\label{eq24}
  \frac{d\orho}{dt}f+\orho f'\frac{d\vphi}{dt}=-\ou_x\orho f
  -\frac32\orho^{1/2}\orho_x\int^{\vphi}\cos\vphi\cdot f'd\vphi
  -\orho^{3/2}\cos\vphi\cdot f'\left.\frac{\prt \vphi}{\prt x}\right|_p.
\end{equation}
Differentiation of Eq.~(\ref{eq22}) with respect to $x$ for $p=\mathrm{const}$ gives
$0=\orho_xf+\orho f'\left.\frac{\prt \vphi}{\prt x}\right|_p$, consequently
\begin{equation}\nonumber
  \left.\frac{\prt \vphi}{\prt x}\right|_p=-\frac{\orho_x}{\orho}\cdot\frac{f}{f'}.
\end{equation}
We calculate the derivative $d\orho/dt$ along the soliton's path:
\begin{equation}\nonumber
  \frac{d\orho}{dt}=\rho_t+V\orho_x=-(\orho\,\ou)_x+(\ou+\sqrt{\orho}\cos\vphi)\orho_x.
\end{equation}
Substitution of these formulas and of Eq.~(\ref{eq19}) into Eq.~(\ref{eq24}) gives after
evident simplifications the equation for $f(\vphi)$:
\begin{equation}\label{eq25}
  f'\sin\vphi=3\int^{\vphi}\cos\vphi f'd\vphi.
\end{equation}
It is readily solved and we obtain
\begin{equation}\label{eq26}
  f=C(\vphi-\sin\vphi\cos\vphi),
\end{equation}
where $C$ is the integration constant (another integration constant can be included into
definition of $\vphi$). This constant factor corresponds to the invariance of the Hamilton
equations with respect to multiplication of $H$ and $p$ by the same constant. For comparison
with earlier works, where the energy unit was fixed by the choice of the Hamiltonian or
Lagrangian for the NLS equation, we choose the value $C=2$. The angle $\vphi$ can be
excluded by means of Eq.~(\ref{eq20}),
\begin{equation}\label{eq27}
  \cos\vphi=\frac{\dot{x}-\ou}{2\sqrt{\orho}}.
\end{equation}
As a result, we arrive at the expressions for the canonical momentum
\begin{equation}\label{eq28}
   p=2\orho\arccos\frac{\dot{x}-\ou}{\sqrt{\orho}}-2(\dot{x}-\ou)\sqrt{\orho-(\dot{x}-\ou)^2}
\end{equation}
and energy
\begin{equation}\label{eq29}
  \begin{split}
  H=\ou p+\frac43\left[\orho-(\dot{x}-\ou)^2\right]^{3/2}.
  \end{split}
\end{equation}
These expressions coincide exactly with formulas derived in Ref.~\cite{shevchenko-88}
for the Bose-Einstein condensate described by the Gross-Pitaevskii equation.
In this derivation the contribution of the counterflow into the momentum was taken into
account and for correct definition of the energy the grand canonical ensemble was used,
so that formation of a soliton did not change the number of particles in the condensate.
As we see, in our derivation based on Stokes' reasoning \cite{stokes}, permitting one to
continue formulas (\ref{eq9}) and (\ref{eq13}) into the soliton region, these effects
are taken into account by default.

Dependence of the Hamiltonian (\ref{eq29}) on the canonical momentum (\ref{eq28}) is
only defined in a parametric way and it hardly can be obtained in an explicit form.
Instead, it is convenient to transform the Hamilton equations to the Newton-like
equation for the variable $x(t)$. This transformation has already been discussed in
Ref.~\cite{ik-22} and we will present here the main results with some additional comments.

\section{Newton equation}

Differentiation of Eq.~(\ref{eq28}) with respect to time gives for the left-hand side of the
Hamilton equation (\ref{eq23}) the expression
\begin{equation}\label{eq30}
\begin{split}
  \frac{dp}{dt}=-4(\ddot{x}-\ou_x\dot{x}-\ou_t)\sqrt{\orho-(\dot{x}-\ou)^2}
  + 2(\orho_x\dot{x}+\orho_t)\arccos\frac{\dot{x}-\ou}{\sqrt{\orho}}.
  \end{split}
\end{equation}
In Hamiltonian mechanics we have $\dot{x}=\dot{x}(x,p)$, so the derivative in the right-hand side
of Eq.~(\ref{eq23}) is equal to
\begin{equation}\label{eq31}
\begin{split}
  \left.\frac{\prt H}{\prt x}\right|_{p}=
  \left.\frac{\prt H^{(0)}}{\prt\orho}\right|_{\dot{x},u}\orho_x
  + \left.\frac{\prt H^{(0)}}{\prt \ou}\right|_{\dot{x},\orho}\ou_x
  + \left.\frac{\prt H^{(0)}}{\prt \dot{x}}\right|_{\orho,\ou}\cdot
  \left.\frac{\prt \dot{x}}{\prt x}\right|_p+\ou_xp,
  \end{split}
\end{equation}
where $H^{(0)}=\frac43[\orho-(\dot{x}-\ou)^2]^{3/2}$. To find the derivative
$\left.\frac{\prt \dot{x}}{\prt x}\right|_p$, we differentiate Eq.~(\ref{eq28})
with respect to $x$ for constant $p$ and obtain
\begin{equation}\label{eq32}
  \left.\frac{\prt\dot{x}}{\prt x}\right|_{p}=\ou_x+
  \frac{\orho_x}{2\sqrt{\orho-(\dot{x}-\ou)^2}}\arccos\frac{\dot{x}-\ou}{\sqrt{\orho}}.
\end{equation}
Elementary calculation of the other derivatives and substitution of these expressions
into Eq.~(\ref{eq30}) yields after simple transformations
\begin{equation}\label{eq33}
\begin{split}
  2\ddot{x}=\orho_x+(\ou+\dot{x})\ou_x+2\ou_t
  +\frac{\orho_t+(\orho\,\ou)_x}{\sqrt{\orho-(\dot{x}-\ou)^2}}\arccos\frac{\dot{x}-\ou}{\sqrt{\orho}}.
  \end{split}
\end{equation}
This equation is derived under supposition that the external potential $U$ does not
deform the soliton's profile, so that in calculations of the energy and the momentum
of the soliton one can use the local values of the background variables
$\orho=\orho(x,t)$, $\ou=\ou(x,t)$, which are solutions of the dispersionless
equations (\ref{eq10}) with $U=0$. According to Eq.~(\ref{eq33}), soliton's dynamics
is governed by the gradients of the background variables which change at characteristic
length much greater than the soliton's width. If for some reason these gradients are
very small (for example, $\orho=\mathrm{const}, \ou=\mathrm{const}$ at the initial
moment of time), then the soliton's dynamics is determined by the $x$-dependent
contribution of the external potential into the soliton's energy. We will assume here
that the gradients of $\orho$ and $\ou$ are large enough and the external potential only
affects the soliton's dynamics via these gradients obeying the dispersionless equations
\begin{equation}\label{eq33b}
	\begin{split}
		\orho_t+(\orho\, \ou)_x=0,\qquad
		\ou_t+\ou\,\ou_x+\orho_x=-U_x.
	\end{split}
\end{equation}
If we exclude $\orho_t$ and $\ou_t$ from Eq.~(\ref{eq33}) with the use of these equations,
we get
\begin{equation}\label{eq34}
  2\ddot{x}=-2U_x-\orho_x+(\dot{x}-\ou)\ou_x
\end{equation}
or
\begin{equation}\label{eq35}
  2\ddot{x}=-U_x+\ou_t+\ou_x\dot{x}=-U_x+\dot{\ou}.
\end{equation}
Eq.~(\ref{eq35}) was obtained in Ref.~\cite{ba-2000} in framework of singular perturbation
theory and this confirms validity of our suppositions.

If $U=0$, then Eq.~(\ref{eq35}) can be integrated and we return to Eq.~(\ref{eq15}).
If the dispersionless flow is stationary,
$\orho=\orho(x)$, $\ou=\ou(x)$, then the Newton equation
\begin{equation}\label{eq36}
  2\ddot{x}=-U_x+u_x\dot{x}
\end{equation}
has the energy integral $H(x,\dot{x})=\mathrm{const}$, where $H(x,\dot{x})$ is given by
Eq.~(\ref{eq29}), which would be difficult to obtain by other methods. In particular,
for the background at rest with $\ou=0$ we see from Eq.~(\ref{eq36}) that soliton's motion
is equivalent to motion of a classical particle with mass $m=2$ in the potential $U(x)$.
Naturally, we reproduce
at once the known results \cite{ba-2000,kp-04} about the frequency $\om_0/\sqrt{2}$ of
the dark soliton oscillations in the condensate confined in a trap with the potential
$U(x)=\om_0^2x^2/2$. Comparison of the Hamiltonian theory of dark soliton motion with
exact numerical solutions of Eq.~(\ref{eq1}) for various stationary and non-stationary
situations can be found in Ref.~\cite{ik-22}.

\section{Generalization on other equations of AKNS scheme}

The theory presented above corresponds to the particular form of the dispersion relation
(\ref{eq8}) for linear waves and the dispersionless limit (\ref{eq10}) of the NLS equation.
But one can notice that many equations of the AKNS scheme can be transformed to the same
formulas for the dispersion relation and the dispersionless limit equations by means of a 
simple change of variables. In particular, all the equations considered in Ref.~\cite{ks-23b} 
admit such a change of variables.

Let the system under consideration be described by two variables $u_1,u_2$. Then in the
dispersionless limit the equations for the background variables $\ou_1,\ou_2$ can be
written in the matrix form
\begin{equation}\label{ad1}
  \bu_t+\mathbb{A}\bu_x=0,\qquad \bu=\left(
                                       \begin{array}{c}
                                         \ou_1 \\
                                         \ou_2 \\
                                       \end{array}
                                     \right),
  \qquad \mathbb{A}=\left(
                      \begin{array}{cc}
                        a_{11} & a_{12} \\
                        a_{21} & a_{22} \\
                      \end{array}
                    \right).
\end{equation}
The characteristic velocities $v_{1,2}$ are defined by the equation
\begin{equation}\label{ad2}
  \left|v\mathbb{I}-\mathbb{A}\right|=v^2-\mathrm{tr}\,\mathbb{A}\cdot v
  +\mathrm{det}\,\mathbb{A}=0
\end{equation}
and they are equal to
\begin{equation}\label{ad3}
  v_{1,2}=\frac12\mathrm{tr}\,\mathbb{A}\pm
  \sqrt{\frac14(\mathrm{tr}\,\mathbb{A})^2-\mathrm{det}\mathbb{A}}.
\end{equation}
Comparison with the formula
\begin{equation}\label{ad4}
  v_{1,2}=\left.\frac{\om}{k}\right|_{k\to0}=\ou\pm\sqrt{\overline{\rho}}
\end{equation}
for the characteristic velocities of the NLS dispersionless limit (see Eq.~(\ref{eq8}))
shows that they can be made identical if we connect them by the relationships
\begin{equation}\label{ad5}
  \ou=\frac12\mathrm{tr}\,\mathbb{A},\qquad
  \overline{\rho}=\frac14(\mathrm{tr}\,\mathbb{A})^2-\mathrm{det}\mathbb{A}.
\end{equation}
If the dispersion relation for linearized system under consideration also
has the form
\begin{equation}\label{ad6}
  \om=k\left(\frac12\mathrm{tr}\,\mathbb{A}\pm
  \sqrt{\frac14(\mathrm{tr}\,\mathbb{A})^2-\mathrm{det}\mathbb{A}+\frac{k^2}{4}}\right),
\end{equation}
then the expressions for the canonical momentum and the soliton's Hamiltonian
are obtained from Eqs.~(\ref{eq28}), (\ref{eq29}) by means of the replacements (\ref{ad5}).
Particular cases of the replacements (\ref{ad5}) have already been used in the literature
(see, e.g., Ref.~\cite{ovs-79}), and in our case they are complemented by the condition
of their compatibility with transformation of the dispersion relation (\ref{ad6}). This means
that our approach can be applied to a wide class of completely integrable equations.
The concrete examples will be considered separately together with appropriate applications.

\section{Conclusion}

The approach suggested in this paper is based on the supposition that the carrier wave
number of linear wave packet propagating along non-uniform evolving background can be
expressed in terms of only background variables (see Eq.~(\ref{eq13})). Examples
considered in Refs.~\cite{sk-23,kamch-23,ks-23b} demonstrate that this supposition
is justified for equations which are completely integrable in framework of the
AKNS scheme and the corresponding formulas turn out to be related with the quasi-classical
limit of Lax pairs. If the formulas analogous to Eq.~(\ref{eq13}) are found, then,
following Stokes \cite{stokes}, we can write a similar formula for the inverse half-width
of a soliton and express its velocity in terms of the background variables (see
Eq.~(\ref{eq15})). The resulting equation is enough for finding the soliton's path
along known background flow if the external potential is absent. If the background
variables depend of the external potential, then, under assumption that the soliton's
dynamics is determined by the gradients of these variables rather than by direct
contribution of the potential to the soliton's energy, we find the Hamilton equations
for soliton's dynamics and the corresponding Newton equation. We demonstrated that
the suggested method can be applied to other equations integrable in the AKNS scheme.

\begin{acknowledgments}
I am grateful to E.~A.~Kuznetsov, N.~Pavloff and D.~V.~Shaykin for useful discussions.
This research is funded by the Advancement of Theoretical Physics and Mathematics ``BASIS''
(section II) and by the research project FFUU-2021-0003 of the Institute of Spectroscopy
of the Russian Academy of Sciences (section III).
\end{acknowledgments}

\end{document}